\documentclass[a4paper,fleqn]{cas-dc}
\usepackage[numbers]{natbib}
\usepackage{siunitx}
\usepackage{braket}
\usepackage{threeparttable}
\usepackage{color}
\usepackage{ulem}

\begin{document}

\let\WriteBookmarks\relax
\def\floatpagepagefraction{1}
\def\textpagefraction{.001}
\shorttitle{Understanding one-body losses in magnetically trapped metastable europium atoms}
\shortauthors{Hiroki Matsui et~al.}

\title [mode = title]{Understanding one-body losses in magnetically trapped metastable europium atoms}                      

\author[1]{Hiroki Matsui}
\author[1]{Yuki Miyazawa}
\author[2]{Ryotaro Inoue}
\author[1,2]{Mikio Kozuma}

\address[1]{Department of Physics, Tokyo Institute of Technology, 2-12-1 O-Okayama, Meguro, Tokyo 152-8550, Japan}

\address[2]{Institute of Innovative Research, Tokyo Institute of Technology, 4259 Nagatsuta, Midori, Yokohama, Kanagawa 226-8503, Japan}

\begin{abstract}
	We report the measurement of one-body loss rates for magnetically trapped metastable europium atoms and the study of their loss mechanism. 
	The loss of atoms observed in a magneto-optical trap is not fully understood because of the indivisibility of the loss regarding optical pumpings due to the presence of cooling laser beams. 
	We magnetically trapped the atoms by directly loading from the magneto-optical trap and observed almost identical one-body loss rates of approximately \SI{2.6}{s^{-1}} for two isotopes: $\mathrm{{}^{151}Eu}$ and $\mathrm{{}^{153}Eu}$.
	Our rate-equation-based model, combined with loss rate measurements carried out with repumpers, shows that blackbody radiation at room temperature drives E1 transitions and induces the observed losses.
\end{abstract}




\maketitle

\section{Introduction}
Quantum degenerate atomic gases are proven platforms for quantum simulations of many-body interacting systems \cite{bloch2012,georgescu2014}.
The degenerate gas of magnetic lanthanides \cite{lu2011,aikawa2012,davletov2020}, which features large mass numbers ($>$$100$), large magnetic dipole moments ($\gtrsim 4\mu_\mathrm{B}$), and large orbital angular momenta ($L\geq 3$), has paved the way for closer studies on quantum few- and many-body systems with anisotropic interactions such as the formation of supersolids\cite{bottcher2019, chomaz2019, tanzi2019a}, the observation of roton modes \cite{chomaz2018, petter2019}, and the emergence of chaotic behavior in Fano-Feshbach resonances \cite{frisch2014,maier2015,khlebnikov2019}.

\begin{figure}
	\centering
	\includegraphics[width=1.0\linewidth]{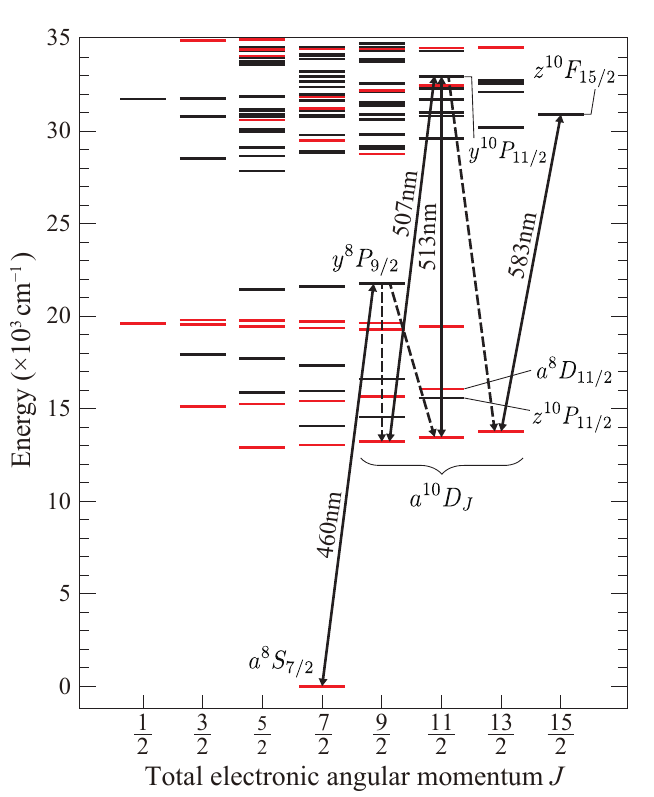}
	\caption{
	Energy level structure of Eu below 35000 $\text{cm}^\text{-1}$ \protect\cite{martin1978,hartog2002}. 
	Odd parity states are shown in gray (red online).
	Arrows indicate transitions utilized in the preparation of cold Eu* atoms ($\text{a}^\text{10}\text{D}_\text{13/2}$), along with their wavelengths.
	}
	\label{FIG:1}
\end{figure}

Magneto-optical trapping (MOT) followed by evaporative cooling in a magnetic \cite{anderson1995,ketterle1996} or optical \cite{barrett2001} trap is a standard scheme for achieving cooling to quantum degeneracy. 
In our previous work, we demonstrated MOT of metastable europium atoms \cite{inoue2018}.
Europium (Eu), also a magnetic lanthanide (with a magnetic dipole moment of $7\mu_\mathrm{B}$ in its electronic ground state), features a nonzero nuclear spin $I=5/2$ and thus has a hyperfine structure in both bosonic isotopes, $\mathrm{{}^{151}Eu}$ and $\mathrm{{}^{153}Eu}$. 
Moderate spacings in its hyperfine structure ($\leq$\SI{121}{MHz}\cite{sandars1960}), in principle, enable control of the contact interactions using microwaves \cite{papoular2010} in the frequency range within the spacings, providing undemanding access to dipolar spinor Bose gases, which are expected to exhibit novel quantum phases \cite{kawaguchi2012}.
Toward its realization, MOT for $\mathrm{{}^{151}Eu}$ was first demonstrated for the metastable state $a^{10}D_{13/2}$ because of the lack of a suitable transition for the laser cooling in the ground state \cite{miyazawa2017a} (see Fig.~\ref{FIG:1}).
Losses observed in the MOT for metastable europium (Eu*) include two-body and one-body losses, as in typical MOTs.
The observed two-body loss rate coefficient $\beta\lesssim 1\times10^{-10}\,$\si{cm^3/s} is comparable to that of metastable noble gases in the absence of resonant lights \cite{vassen2012} and magnetic lanthanides in narrow-line MOTs \cite{frisch2012,maier2014,tsyganok2018}.
For the one-body loss rates in Eu* MOT, its strong dependency on the intensity of cooling laser beams has been observed, and its partial suppression was achieved with pumping laser beams tuned to the $a^{10}D_{9/2}\rightarrow y^{10}P_{11/2}$ and $a^{10}D_{11/2}\rightarrow y^{10}P_{11/2}$ transitions.
These behaviors could be explained by one-body loss paths to the $a^{10}D_{9/2}$ and $a^{10}D_{11/2}$ levels via the excited state $z^{10}F_{15/2}$ of the cooling transition.
Moreover, Eu* has an E1 transition to the $z^{10}P_{11/2}$ level, which could be driven by room-temperature blackbody radiation, and its contribution to the one-body loss was suggested by observation of the MOT.
A qualitative understanding of the one-body loss mechanism has yet to be acquired.

We have magnetically trapped Eu* atoms loaded directly from the MOT and measured one-body loss rates of atoms for both isotopes, thereby investigating the inevitable losses in a room-temperature chamber.
The observations confirm that one-body losses occur without the MOT cooling laser beams, with no significant difference observed between the isotopes.
We also confirm that some of the lost atoms can be optically pumped back to the metastable state $a^{10}D_{13/2}$ or eventually decay to the ground state $a^{8}S_{7/2}$.
A rate-equation-based model consistently explains the observed loss rates by taking into account transitions induced by room-temperature blackbody radiation.

\section{Eu* in a magnetic trap}
Figure~\ref{FIG:1} shows our scheme for the preparation of Eu* in the $a^{10}D_{13/2}\,(F=9)$ state with its cooling transition for the Zeeman slower and MOT \cite{inoue2018}.
Eu atomic vapors are produced in an oven heated to \SI{770}{K}.
Atoms from the oven are optically pumped to various metastable states via the $a^{8}S_{7/2}\,(F=6)\rightarrow y^{8}P_{9/2}\,(F=7)$ transition at \SI{460}{nm} with a natural linewidth of $2\pi\times 27\,\mathrm{MHz}$.
Atoms in two of those states, i.e. $a^{10}D_{9/2}\,(F=7)$ and $a^{10}D_{11/2}\,(F=8)$, are then pumped to the $a^{10}D_{13/2}$ $(F=9)$ state by two laser lights tuned to the $a^{10}D_{9/2}(F=7)\rightarrow y^{10}P_{11/2}(F=8)$ transition at \SI{507}{nm} and the $a^{10}D_{11/2}(F=8)\rightarrow y^{10}P_{11/2}(F=8)$ transition at \SI{513}{nm}.
The Eu* atoms are then decelerated in the Zeeman slower and captured by MOT, both utilizing the \SI{583}{nm} quasi-cyclic $a^{10}D_{13/2}\,(F=9)\rightarrow z^{10}F_{15/2}\,(F'=10)$ transition with a natural linewidth of $2\pi\times 8.2\mathrm{MHz}$.
The Zeeman-slowing and MOT-cooling beams are red-detuned by \SI{244}{MHz} and \SI{17.5}{MHz}, respectively, from the transition resonance.
The MOT is composed of a quadrupole magnetic field with a radial field gradient of \SI{25}{mT/m} and six cooling laser beams with an intensity of \SI{4.0}{mW/cm^2} in total. 
The hyperfine repumping laser beams tuned to the $F=8\rightarrow F'=9
$ transition propagate together with all six laser beams.
The trap typically contains \num{2.4(3)e+7} atoms for $\mathrm{{}^{151}Eu}$ and \num{3.7(4)e+7} atoms for $\mathrm{{}^{153}Eu}$.
The trapped atoms possess a two-temperature momentum distribution, as described in our previous work\cite{inoue2018}. 
In this work, the distribution is characterized by temperatures of \SI{0.03(1)}{mK} and \SI{0.53(8)}{mK}, each part containing $\sim$\SI{50}{\percent} of the total atoms.

Before loading the collected Eu* atoms to the magnetic trap (MT), we spin-polarize them to improve the transfer efficiency: the $F=9$ hyperfine level has 19 magnetic sublevels.
Our spin-polarization procedure starts by turning off all the laser beams and the magnetic field gradient used for the MOT.
A bias magnetic field is then applied, and the pumping light, which consists of $\sigma^+$-polarized two-color lasers tuned to the $F=9\rightarrow F'=9$ and the $F=8\rightarrow F'=9$ transitions, is introduced to the atoms and retroreflected to avoid directional momentum kicks during the \SI{200}{\micro s}-long polarizing step.
We then load the atoms into the MT with a quadrupole magnetic field by ramping up the field within \SI{1}{ms}.
The radial field gradient of \SI{185}{mT\per m} applied in the MT is optimized to avoid excess heating and dilution of the higher-temperature components and thus confine all the atoms with minimal spatial deformation.
This field gradient corresponds to a trap depth of \SI{90}{mK}.
A mechanical shutter blocks stray light near-resonant to the $a^{10}D_{13/2}\rightarrow z^{10}F_{15/2}$ transition during the MT, thereby preventing spin-flips.

The number of Eu* atoms in the trap and their distribution are evaluated by an absorption imaging technique utilizing the $F=9\rightarrow F'=10$ transition after the magnetic field gradient is turned off.
We found that the transfer efficiency into the MT is greater than \SI{96}{\%} for both isotopes.
The number of Eu* atoms in the trap exponentially decays over time, indicating that there exist one-body loss mechanisms that function even in the absence of MOT cooling laser beams.
There are no significant differences in the observed one-body loss rates between the two isotopes: \SI{2.60(2)}{s^{-1}} for ${}^{151}\mathrm{Eu}$ and \SI{2.68(4)}{s^{-1}} for ${}^{153}\mathrm{Eu}$.
The atoms in the MT exhibit breathing oscillations in the spatial distribution, indicating imperfect matching of the distribution in the MOT and the MT.
These initial oscillations dephase within \SI{80}{ms} due to anharmonicity of the quadrupole MT and are no longer apparent for the longer holding times.
Although the atoms are not in thermal equilibrium even at a holding time of \SI{1}{s}, the cloud sizes are characterized by half widths at half maximum of the atomic density distribution, with values of \SI{0.28(3)}{mm} in the radial directions and \SI{0.09(1)}{mm} in the axial direction for both isotopes.
During the observations of one-body loss after oscillation dephasing, the cloud sizes have remained constant.
At a holding time of \SI{100}{ms}, the peak number densities of atoms at the trap center are \SI{0.6(2)e+12}{cm^{-3}} for $\mathrm{{}^{151}Eu}$ and \SI{1.0(3)e+12}{cm^{-3}} for $\mathrm{{}^{153}Eu}$.

\begin{figure}[tbp]
	\centering
	\includegraphics[width=1.0\linewidth]{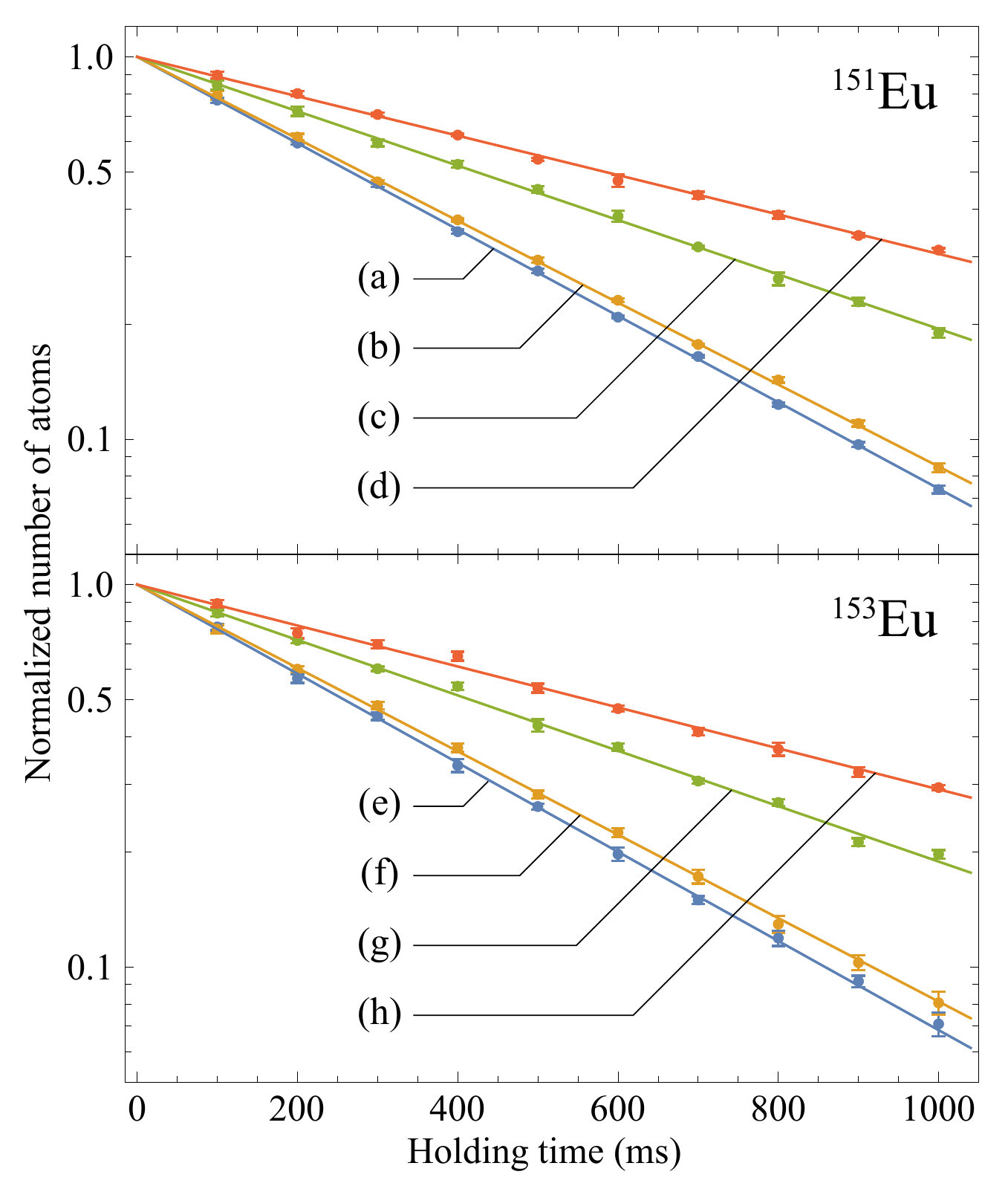}
	\caption{
	Normalized number of ${}^\text{151}\text{Eu}$ (upper) and ${}^\text{153}\text{Eu}$ (lower) atoms as a function of the MT holding time for atoms in the $\text{a}^\text{10}\text{D}_\text{13/2}$ state under different conditions: (a), (e) without repumpers; (b), (f) with the $\text{a}^\text{10}\text{D}_\text{9/2}\rightarrow \text{y}^\text{10}\text{P}_\text{11/2}$ repumper; (c), (g) with the $\text{a}^\text{10}\text{D}_\text{11/2}\rightarrow \text{y}^\text{10}\text{P}_\text{11/2}$ repumper; and (d), (h) with both repumpers.
	Fits by the exponential function (solid lines) give the decay rates listed in Tab.~\ref{TBL:1}.
	Each error bar indicates the standard error estimated from four independent measurements.
	}
	\label{FIG:2}
\end{figure}

Our previous results on the MOT of Eu* suggested that there are one-body radiative loss channels related to the\linebreak $a^{10}D_{9/2}$ and $a^{10}D_{11/2}$ levels even in the absence of cooling laser beams.
To examine the contribution of the loss channels to the one-body loss rate, we illuminated the magnetically trapped Eu* atoms with the two pumping laser beams used for preparing Eu* ($a^{10}D_{9/2}(F=7)\rightarrow y^{10}P_{11/2}(F=8)$ and $a^{10}D_{11/2}(F=8)\rightarrow y^{10}P_{11/2}(F=8)$ repumpers, see Fig.~\ref{FIG:1}).
Figure~\ref{FIG:2} shows variations of the number of Eu* atoms with time in the MT under four conditions: in the presence of one, both, or none of the pumping beams.
It is apparent from the figure that the existence of the pumping beams reduces the decay rate.
Additionally, the continuous repumping during MT is essential for suppressing the decay rate, which suggests that the atoms in $a^{10}D_{9/2}$ and $a^{10}D_{11/2}$ decayed to the other levels.

\begin{table*}[tbp]
	\centering
	\begin{threeparttable}
	\caption{
		Experimentally observed and theoretically calculated decay rates in $\text{s}^\text{-1}$ under the four repumping conditions.
		The decay rate without the repumpers ($\text{2.6}\,\text{s}^\text{-1}$) is used to calculate the others.
		The calculated decay rates are shown with two significant digits because the calculation model contains levels with purities less than \SI{90}{\percent} for the bases in the $LS$-coupling scheme.
		See Sec.~\ref{SEC:A} for the calculation details.
		}
	\label{TBL:1}
	\begin{tabular*}{0.7\textwidth}{LCCC}
		\toprule
		 & ${}^\text{151}$Eu & ${}^\text{153}$Eu & Rate-equation model\\
		\midrule
		without repumpers & 2.60(2)\tnote{(a)} & 2.68(4)\tnote{(e)} & (2.6)\\
		with $\text{a}^\text{10}\text{D}_\text{9/2}\rightarrow \text{y}^\text{10}\text{P}_\text{11/2} $ repumper & 2.46(2)\tnote{(b)} & 2.51(4)\tnote{(f)} & 2.4\\
		with $\text{a}^\text{10}\text{D}_\text{11/2}\rightarrow \text{y}^\text{10}\text{P}_\text{11/2}$ repumper & 1.64(3)\tnote{(c)} & 1.67(3)\tnote{(g)} & 1.4\\
		with repumpers & 1.19(2)\tnote{(d)} & 1.23(3)\tnote{(h)} & 1.1\\
		\bottomrule
	\end{tabular*}
	\begin{tablenotes}
		\item[(a)-(h)]{Symbols correspond to those in Fig.~\ref{FIG:2}.}
	\end{tablenotes}
	\end{threeparttable}
\end{table*}

\begin{figure}[b]
	\centering
	\includegraphics[width=1.0\linewidth]{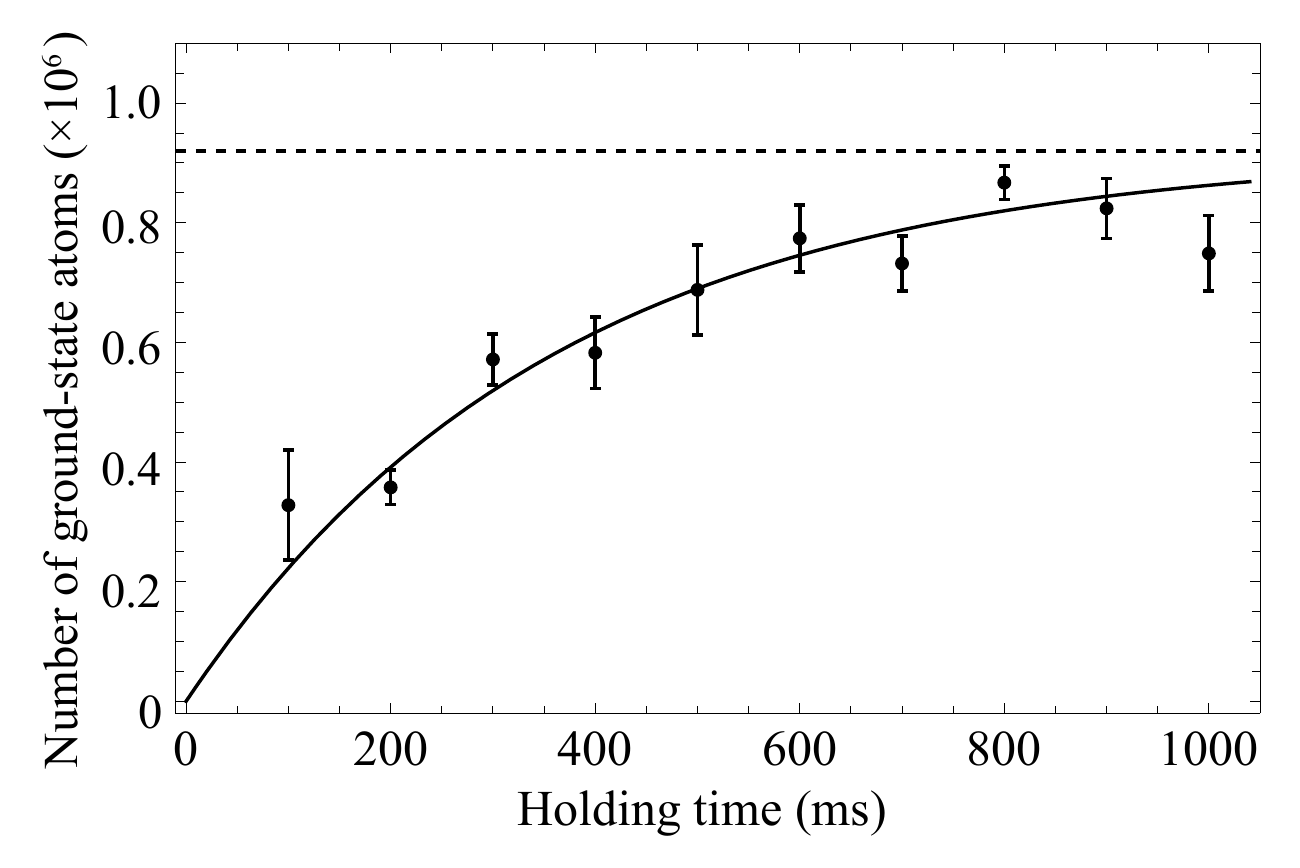}
	\caption{
	Time evolution of the ground-state Eu ($\text{a}^\text{8}\text{S}_\text{7/2}$ $\text{(F=6)}$) atoms in the MT. 
	The solid curve denotes the fit to the data (see text), and its asymptote is indicated by the dotted line.
	}
	\label{FIG:3}
\end{figure}

Figure~\ref{FIG:3} shows the observed spontaneous growth in the ground-state population in the MT for ${}^{151}\mathrm{Eu}$.
Here, the number of trapped atoms is observed by an absorption imaging technique utilizing the $a^{8}S_{7/2}\,(F=6)\rightarrow y^{8}P_{9/2}\,(F'=7)$ transition.
For technical reasons, a reduced number of Eu* atoms (\num{2.4e+6}) are loaded into the MT, resulting in the relatively large errors in Fig.~\ref{FIG:3}.
Assuming a simple case where the numer of ground-state atoms $N_\mathrm{g}(t)$ represents a portion of the number of metastable atoms $\alpha\cdot N_\mathrm{m}(t)$ with the decay rate $\Gamma_\mathrm{mg}$, one can yield the simple model:
\begin{align}
	N_\mathrm{g}(t)=\alpha\cdot N_\mathrm{m}(0)\left(1-\exp(-\Gamma_\mathrm{mg}t)\right).
	\label{EQ:1}
\end{align}
We obtained a metastable to ground-state transfer efficiency $\alpha=0.38(2)$ and a decay rate $\Gamma_\mathrm{mg}=$ \SI{2.8(4)}{s^{-1}} from the curve fitting.
The rate $\Gamma_\mathrm{mg}$ is in good agreement with the decay rate (\SI{2.60(2)}{s^{-1}}) for the metastable atoms without repumpers.
Note that we also assumed a fully polarized sample to evaluate the number of ground-state atoms, though spin relaxation might exist during the decay process.
Therefore, the estimated number of ground-state atoms and the obtained transfer efficiency $\alpha$ include non-negligible systematic errors but represent lower bounds for any possible spin distributions.

\section{Loss mechanisms of Eu*}
\label{SEC:3}
The observed exponential decay in the number of Eu* atoms indicates that one-body loss processes are predominant over collisional loss processes, including magnetic dipole-dipole interactions.
The Majorana spin-flip transition is a one-body process that can result in the loss of an atom and is concerned with a MT, especially one with a quadrupole field.
The Majorana transition rate can be estimated to be on the order of \SI{e-2}{s^{-1}} in our setup considering the quadrupole field with the measured cloud geometry \cite{petrich1995}.
The losses accompanied by these Majorana transitions therefore contribute little to the observed loss rate.
Additionally, since the pressure in the trapping chamber is less than \SI{e-9}{Pa}, collisions with background gasses can be neglected.
The observed reduction in the loss rates under continuous exposure to the repumping laser beams signifies that the Eu* atoms decay to the other levels without gaining enough energy to escape from the trap.
Therefore, it is inferred that the dominant loss process of Eu* is a decay via optical transitions, in which photons take out the energy difference between the initial and the final states.

Eu* has two transitions that could be involved in its loss process: the $a^{10}D_{11/2}\rightarrow a^{10}D_{13/2}$ and $a^{10}D_{13/2}\rightarrow z^{10}P_{11/2}$ transitions.
The former is an M1 transition within the fine structure.
A population inversion occurs in this two-level system at the beginning of the MT.
The upper ($a^{10}D_{13/2}$) and lower ($a^{10}D_{11/2}$) levels have \SI{95}{\%} and \SI{94}{\%} purity for the basis vector represented in each term symbol, respectively \cite{martin1978}.
We thus adopt the pure $LS$-coupling scheme here to estimate the transition probability \cite{condon1935,pasternack1940}.
Considering blackbody radiation from a room-temperature environment, one yields loss rates of \SI{1.2e-3}{s^{-1}} for the spontaneous and \SI{0.3e-3}{s^{-1}} for the stimulated transitions.
The calculated rates are three orders of magnitude smaller than the observation, which indicates that the M1 transition is unlikely to be responsible for the observed loss.
The direct decay via the $a^{8}S_{7/2}\rightarrow a^{10}D_{13/2}$ transition seems more unlikely to be relevant to the loss than the M1 transition for several reasons: the initial and the final states have the same wave function parity, different spin multiplicities, and a large difference in the total angular momentum of $\Delta J = 3$.

The latter candidate, the $a^{10}D_{13/2}\rightarrow z^{10}P_{11/2}$ transition, is an E1 transition.
In a loss process that involves the E1 transition, Eu* atoms are excited to the $z^{10}P_{11/2}\,(F'=8)$ state by blackbody radiation with a wavelength of \SI{5.5}{\micro m} and then decay to the other states.
Though no information on the oscillator strength of the E1 transition is available to the best of our knowledge, we have constructed a simple model that considers the E1 transition and explains the observation well.
We propose the model and discuss the loss mechanism of Eu* in the following part.

To numerically simulate the population dynamics of\linebreak trapped atoms, we consider only E1 transitions, including transitions caused by room-temperature blackbody radiation.
\begin{figure}
	\centering
	\includegraphics[width=1.0\linewidth]{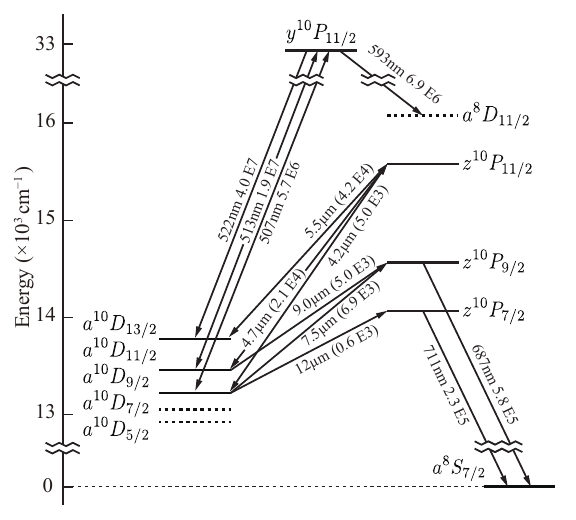}
	\caption{
		Whole picture of the model used to simulate the population dynamics in a magnetic trap.
		Arrows indicate the transitions considered in the calculations along with their wavelengths and the transition probabilities in \si{s^{-1}} \protect\cite{hartog2002,morton2000}. 
		The transition probabilities in parentheses are our estimations.
	}
	\label{FIG:4}
\end{figure}
Since the blackbody radiation is the only driving factor, we ignore optical coherences and create a model composed of rate equations for the populations in the various states.
A rate equation has to be set up separately for each of the magnetic sublevels because atoms can leave an MT flipping their spins to the untrapped states ($m_F\leq 0$) by optical transitions.
All the atoms in a trap are initially populated in the fully stretched state: $a^{10}D_{13/2}\,(F=9;\, m_F=+9)$.
Blackbody radiation excites atoms populated in the $a^{10}D$ levels to the $z^{10}P$ levels.
Atoms excited to the $z^{10}P_{11/2}$ level decay to other levels, which includes the initial and the other hyperfine states of the initial level $a^{10}D_{13/2}$.
In contrast, the $z^{10}P_{9/2}$ and the $z^{10}P_{7/2}$ levels have strong optical transitions to the ground level $a^{8}S_{7/2}$ with transition energy much greater than of the blackbody radiation.
Thus, most of the atoms excited to these two levels decay to the ground level and are no longer involved in the population dynamics.
Therefore, transitions to the $a^{10}D_{7/2}$ and $a^{10}D_{5/2}$ levels are neglected in the model.

We lack sufficient information on the oscillator strengths of transitions from $a^{10}D_J$ levels to $z^{10}P_{J'}$ levels for the $J$ and $J'$ combinations to construct a model.
However, since all the relevant levels have purities greater than \SI{80}{\%} for the respective basis in the $LS$-coupling scheme \cite{martin1978}, we fix the ratios between the oscillator strengths and reduce the number of unknown parameters to only one by assuming the pure $LS$-coupling scheme.
Then, we deduce all the oscillator strengths in question from the observed one-body loss rate (\SI{2.6}{s^{-1}}) by solving the rate equations for blackbody radiation at \SI{300}{K}.
The obtained oscillator strengths give the transition probabilities of \SI{6.9e+3}{s^{-1}} for the $a^{10}D_{9/2}\rightarrow z^{10}P_{9/2}$ and \SI{5.0e+3}{s^{-1}} for the $a^{10}D_{11/2}\rightarrow z^{10}P_{9/2}$ transitions
\footnote[1]{
	Recently, we demonstrated MOT for ground-state Eu atoms for the first time \cite{miyazawa2021}.
	Spectroscopy of the ground-state MOT yields the transition probabilities \SI{7.4(2)e+3}{s^{-1}} and \SI{4.5(1)e+3}{s^{-1}} for the $a^{10}D_{9/2}\rightarrow z^{10}P_{9/2}$ and $a^{10}D_{11/2}\rightarrow z^{10}P_{9/2}$ transitions, respectively.
	The slight deviation of our calculation from this observation may be attributable to adopting the pure $LS$-coupling scheme, as implied by the disagreement in the ratios of the transition probabilities for the two transitions.
}.
The resultant rate equations are provided in Sec. \ref{SEC:A}.

We calculated the time evolution in the ground-state population using the obtained model.
The model yielded a metastable- to ground-state transfer efficiency of $\alpha=0.90$ and a transfer rate of $\Gamma_\mathrm{mg}=2.5\,\mathrm{s^{-1}}$, with a delay of $\delta t=47\,\mathrm{ms}$ in the growth of the ground-state population.
The transfer rate $\Gamma_\mathrm{mg}$ is within the error range of the experimental result ($\Gamma_\mathrm{mg}=2.8(4)\,\mathrm{s^{-1}}$).
The delay $\delta t$ obtained from the calculation reflects the duration of stays in the multiple intermediate states, though it is not clearly visible in Fig.~\ref{FIG:3} since data points are not available for holding times shorter than $100\mathrm{ms}$.
The deviation of the calculated transfer efficiency $\alpha$ from the observation ($\alpha=0.38(2)$) could be attributed to the systematic errors originating from spin relaxations of atoms and the assumption of pure $LS$-coupling in the calculation model.

We also simulated the population dynamics of trapped Eu* under the same condition as in the experiments with the repumping laser beams. 
Figure~\ref{FIG:4} shows the whole system involved in the dynamics.
The oscillator strengths obtained above are applied to the model.
The actual intensities and detunings of the repumping laser beams are used, but their polarization, which atoms feel, is assumed to average out for positions in the trap to simplify the calculation.
In addition, although these beams are coherent light sources, their pump rates are four orders of magnitude smaller than the spontaneous decay rate of the upper-level $y^{10}P_{11/2}$; thus, the coherences do not grow and can be ignored.
Therefore, one can use the rate-equation-based model for the simulation.
Atoms excited by the repumping lights can decay to the $a^{8}D_{11/2}$ level, which is treated as a sink in the model because of having a spin multiplicity different from that of the $a^{10}D$ levels.
The one-body loss rates of Eu* calculated with the model are listed in Tab.~\ref{TBL:1}.
The calculated rates are within \SI{20}{\%} of the observed rates.
The results confirm that the model can reproduce the trends in the observations.
According to the model, the loss rates of the E1 transitions induced by blackbody radiation in a chamber at \SI{200}{K} would be comparable to that of the Majorana losses $(\sim 10^{-2}\,\mathrm{s^{-1}})$ and at \SI{170}{K} that of the spontaneous M1 transitions $(\sim 10^{-3}\,\mathrm{s^{-1}})$.

\section{Conclusions}
We have demonstrated the magnetic trapping of Eu* atoms for both ${}^{151}\mathrm{Eu}$ and ${}^{153}\mathrm{Eu}$ isotopes.
The observation of exponential decay in the number of magnetically trapped atoms confirms that Eu* has a one-body loss mechanism that functions without any laser beams, as implied previously by the loss rate measurements for Eu* MOT.
The observations are successfully explained by a model that attributes the observed one-body losses of Eu* to the E1 transitions driven by room-temperature blackbody radiation.
Losses induced by blackbody radiation are inevitable in a room-temperature chamber; therefore, in the future production of Eu BEC, we encourage pumping Eu* atoms back to the ground state as soon as they are laser-cooled.

\section{Acknowledgments}
This work was supported by JST Grant Numbers JPMJPF2015, JPMJMI17A3, and JSPS KAKENHI Grant Numbers JP16K13856, JP17J06179, and JP20J21364.
H. M. also acknowledges the financial support from Advanced Research Center for Quantum Physics and Nanoscience, Tokyo Institute of Technology. 

\numberwithin{equation}{section}

\appendix
\section{Rate equation model}
\label{SEC:A}

Our model can be used to calculate populations in the levels designated with solid lines in Fig.~\ref{FIG:4}.
Each level consists of six hyperfine levels reflecting a nuclear spin of $I=5/2$, and each hyperfine level comprises magnetic sublevels.
These sublevels have to be treated individually in the calculation because atoms that have transitioned to the untrapped states ($m_F\leq 0$) escape from the MT and thus have to be removed from the calculation.
As a result, our model is composed of 516 magnetic sublevels in total.
We hereinafter uniquely assign a label for each magnetic sublevel defined along quantization axes parallel to local magnetic fields and define $U$ as the set of all 516 labels.
Since spontaneous transitions are unidirectional, the upper and lower states have to be handled separately for each transition.
For each label $i\in U$, we define a collection of the upper-state labels as $U_{\mathrm{upper,\,}i}$ and a collection of the lower-state labels as $U_{\mathrm{lower,\,}i}$ with respect to the state labeled as $i$.

The E1 transition is the only transition mechanism considered in our model.
Spontaneous transition rates $a_{\mathrm{spt},ij}$ ($i\in U,\, j\in U_{\mathrm{lower,\,}i}$) are obtained from the transition probabilities in Fig.~\ref{FIG:4} using the Wigner-Eckart theorem under the assumption of pure $LS$-coupling.
Similarly, stimulated transition rates for repumping lights $a_{\mathrm{rep},ij}\, (i,j\in U)$ are obtained from the excitation rates: \SI{30}{s^{-1}} for the $a^{10}D_{9/2}(F=7)\rightarrow y^{10}P_{11/2}(F=8)$ repumper and \SI{2E+4}{s^{-1}} for the $a^{10}D_{11/2}(F=8)\rightarrow y^{10}P_{11/2}(F=8)$ repumper, each estimated from the experimental parameters.
Here, the energy density of repumping lights is assumed to be equally divided into the three polarizations as a result of averaging over the space around the trap center.
Note that the accuracies of an order in the excitation rates are sufficient to calculate the decay rates with errors of less than \SI{10}{\percent}.
The accuracies of the calculated decay rates are rather limited by the fact that the calculation model contains levels with purities less than \SI{90}{\percent} for the bases in the $LS$-coupling scheme, which causes deviations of transition probabilities in our model from those of actual atoms.
The stimulated transition rates for blackbody radiation $a_{\mathrm{BBR},ij}\, (i,j\in U)$ at room temperature $T=300\mathrm{\, K}$ are given as
\begin{align}
	a_{\mathrm{BBR},ij}=a_{\mathrm{spt},ij}\frac{1}{e^{h\nu_{ij}/k_\mathrm{B}T}-1}\, ,
	\label{EQ:a-1}
\end{align}
where $\nu_{ij}$ is the transition frequency.
By summing the contributions from blackbody radiation and the repumping lights, we obtain the total stimulated transition rate between the states with labels $i$ and $j$ as
\begin{align}
	a_{\mathrm{stm},ij}=a_{\mathrm{BBR},ij}+a_{\mathrm{rep},ij}\, .
	\label{EQ:a-3}
\end{align}

Escapes from the population dynamics via a state with a label $i\in U$ compose a loss rate $\gamma_i$ in our model.
Transitions to the untrapped states ($m_F\leq 0$) can result in losses of atoms in MTs.
However, excitations to the untrapped states do not immediately result in losses because the lifetimes of the upper states are not sufficient for atoms to escape from the MT.
Therefore, we incorporate transitions to the lower untrapped states, including stimulated and spontaneous ones, in the loss rates $\gamma_i \,(i\in U)$.
Additionally, we also incorporate the decay $y^{10}P_{11/2}\rightarrow a^8D_{11/2}$ into the loss rates $\gamma_i$ for the reasons mentioned in Sec. \ref{SEC:3}.

According to the above, the rate equations for the populations $N_i\,(i\in U)$ read as
\begin{align}
	&\frac{dN_i}{dt}=\sum_{j\in U_{\mathrm{upper,\,}i}}
	\left\{N_j (a_{\mathrm{stm},ji}+a_{\mathrm{spt},ji})-N_i\, a_{\mathrm{stm},ij} \right\}\nonumber\\
	&\quad+\sum_{j\in U_{\mathrm{lower,\,}i}}
	\left\{N_j a_{\mathrm{stm},ji}-N_i(a_{\mathrm{stm},ij}+a_{\mathrm{spt},ij}+\gamma_i)\right\}\, .
	\label{EQ:a-4}
\end{align}


\bibliographystyle{elsarticle-num}

\bibliography{matsui_paper1}

\end{document}